# Embracing defects and disorder in magnetic nanoparticles

*Aidin Lak,*[*a] *Sabrina Disch*[*b] *and Philipp Bender*[*c]

[a]Department of Physics and Center for NanoScience, LMU Munich, Amalienstr. 54, 80799 Munich, Germany.
[b]Department für Chemie, Universität zu Köln, Luxemburger Strasse 116, 50939 Köln, Germany.
[c]Department of Physics and Materials Science, University of Luxembourg,162A avenue de la Faïencerie, L-1511 Luxembourg, Grand Duchy of Luxembourg.

E-mails: lak.aidin@lmu.de, sabrina.disch@uni-koeln.de, philipp.bender@uni.lu

*All three authors contributed equally to the manuscript.*

**Abstract**

Iron oxide nanoparticles have tremendous scientific and technological potential in a broad range of technologies, from energy applications to biomedicine. To improve their performance, single-crystalline and defect-free nanoparticles have thus far been aspired. However, in several recent studies defect-rich nanoparticles outperform their defect-free counterparts in magnetic hyperthermia and magnetic particle imaging. Here, an overview on the state-of-the-art of design and characterization of defects and resulting spin disorder in magnetic nanoparticles is presented with a focus on iron oxide nanoparticles. The beneficial impact of defects and disorder on intracellular magnetic hyperthermia performance of magnetic nanoparticles for drug delivery and cancer therapy is emphasized. Defect-engineering in iron oxide nanoparticles emerges to become an alternative approach to tailor their magnetic properties for biomedicine, as it is already common practice in established systems such as semiconductors and emerging fields including perovskite solar cells. Finally, perspectives and thoughts are given on how to deliberately induce defects in iron oxide nanoparticles and their potential implications for magnetic tracers to monitor cell therapy and immunotherapy by magnetic particle imaging.

**Keywords:** iron oxide nanoparticles, magnetic hyperthermia, defect-engineering, design-by-disorder, magnetic spin disorder

# 1. Introduction

Despite their common negative connotation, *defects* and *disorder* can be highly desirable in nature and materials design[1]. Colorful examples for nature's *design-by-disorder* are butterfly wings, whose vibrant colors originate from slightly disordered arrangements of photonic crystals[2], and white beetle scales, whose exceptional whiteness is caused by a very high degree of disorder[3]. Having learned from nature, mimicking its powerful *design-by-disorder* strategy has become an emerging field of research in materials science[4]. This design method has led to novel applications e.g. within the field of disordered photonics[5]. Nowadays, defect-engineering is commonly used in designing semiconductors[6], perovskite solar cells[7], metals[8], multiferroics[9], nanocomposites[10], and carbon materials[11,12] to tailor functional properties. Recent years have witnessed ground-breaking work in this area[13–16], focusing exclusively on exploiting structural disorder in various functional materials such as metal-organic frameworks to optimize specific material characteristics. Notably, the electronic structure of condensed matter is significantly affected by intrinsic defects such as vacancies, impurities, and dislocations[17], which in turn modify macroscopic material characteristics, including optical[18] and magnetic properties e.g. by domain wall pinning[19].

However, for magnetic materials changing our negative perception about defects has had a much slower pace than other fields. Recently, the generic relevance of the *defect-induced* Dzyaloshinskii–Moriya interaction (DMI) for the spin microstructure of defect-rich ferromagnets was verified[20], suggesting a potential for creation of local chiral spin textures, i.e. skyrmions[21], in all kinds of disordered magnetic materials. Magnetic materials offer wide-spread novel technological potential[22], with magnetic nanoparticles[23], especially iron oxide nanoparticles, being indispensable candidates for varieties of biomedical applications[24], including magnetic hyperthermia[25], and magnetic particle and resonance imaging (MPI[26], MRI[27]).

To assess and optimize the performance of magnetic nanoparticles, it is crucial to consider their structure and magnetism on a variety of length scales, ranging from the atomistic to the macroscopic regime (**Figure 1**). While considering magnetic nanoparticles as mesoscale dipoles is sufficient to understand the qualitative principle of magnetic hyperthermia and MPI, the internal MNP structure directly impacts magnetic properties and, hence, magnetic hyperthermia and MPI performance. Due to their large surface to volume ratio, magnetic nanoparticles are particularly prone to structural and compositional defects,[28] and thus synthesis of defect-free magnetic nanoparticles remains highly challenging.[29] From very early aqueous synthesis of iron oxide nanoparticle suspensions through a co-precipitation method,[30] there have been enormous efforts on controlling physicochemical properties, often with the goal to prepare single-crystalline iron oxide nanoparticles.[31] To date, the synthesis of defect-free iron oxide nanoparticles remains the golden standard in colloidal chemistry.

Recently, however, it was demonstrated that in case of magnetic hyperthermia (and also MPI, MRI) defect-rich iron oxide nanoparticles can actually outperform their defect-free counterparts.[32–39] Motivated by these striking results, here, we will remove the common stigma of defects in iron oxide nanoparticles. At first, we discuss recent studies on positive impacts of defects and disorder on magnetic hyperthermia performance of iron oxide nanoparticles. Next, we review the magnetic spin disorder in iron oxide nanoparticles associated with defects and disorder. Finally, we present an overview about the state-of-the-art on how to incorporate and induce defects in iron oxide nanoparticles.

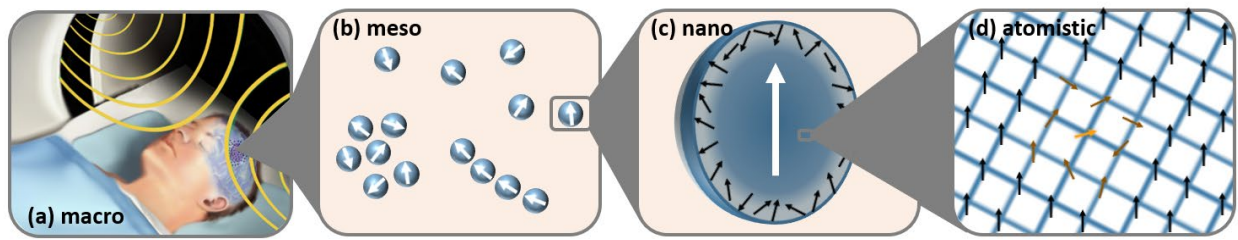

**Figure 1. Multiple length scales in magnetic nanoparticles.** (a) The macroscopic application (here: magnetic hyperthermia for the treatment of glioblastoma) can be understood qualitatively by considering magnetic nanoparticles on the mesoscale as magnetic dipoles (b) of the respective bulk material with nanosized dimensions. Quantitatively, magnetic nanoparticles performance is critically affected by spin disorder on both (c) the nanoscale (i.e. its distribution within the particle from surface into the particle interior) and (d) the atomic scale (e.g. site-defect induced spin disorder). As shown e.g. in[63], defects in the crystalline structure cause atomistic spin disorder, that can critically modify the magnetic properties of nanoparticles and hence influence the macroscopic ensemble properties and therefore their magnetic hyperthermia performance. Panel (a) reproduced from[45].

## 2. Iron oxide nanoparticles in biomedicine

### 2.1. Designing the ideal iron oxide nanoparticles for magnetic hyperthermia

Iron oxide nanoparticles transduce magnetic energy to heat through magnetic losses when exposed to external alternating magnetic fields. They can therefore serve as remotely activated, nanometric heat sources[40] to eradicate cancer cells after being intracellularly engulfed.[25,41] Remarkably, the concept of exploiting iron oxide nanoparticles as local heat sources has been recently expanded to other technologies such as catalysis,[42] water electrolysis,[43] and local polymerization.[44]

Magnetic hyperthermia is especially attractive for the treatment of inoperable tumors such as glioblastoma (illustrated in Figure 1a), the most common and a highly aggressive brain cancer.[45] Although magnetic hyperthermia is not yet a routine treatment, clinical trials are ongoing. With the clinical application of magnetic hyperthermia in mind, the following restrictions need to be considered regarding particle design to optimize the heating:[46] (1) focus on biocompatibility[47] and biodegradation: iron oxide nanoparticles are hitherto the only clinically approved heat transducers to be administered to humans.[48] (2) focus on Néel relaxation: iron oxide nanoparticles have to preserve their heating performance when intracellularly immobilized, as in the cellular milieu a physical rotation of nanoparticles, i.e. Brownian relaxation, is mostly inhibited.[49–51] (3) focus on high-frequency and low-amplitude alternating fields: intracellularly immobilized iron oxide nanoparticles should induce hyperthermic effects, i.e. rising temperature to ~42°C, but also comply with the biological safety limits.[52] Currently, most magnetic hyperthermia devices approved for clinical trials apply magnetic fields within a frequency range of 0.05-1.2 MHz and an amplitude range of 0-5 kA/m.[53]

The macroscopic heating power of an iron oxide nanoparticle ensemble is usually given by the specific absorption rate $SAR = f \cdot S/c$, where $S = \mu_0 \oint M(H)dH$ is the area of alternating field hysteresis loops, and $c$ is the weight concentration of the magnetic material.[54] To better compare results from different experimental setups, the intrinsic loss parameter $ILP$ was introduced, which is given by $ILP = SAR/(\mu_0 H_0^2 f)$.[55,56] Within the framework of the linear response theory, that is at first approximation valid in case of low amplitudes, the $ILP$ is given by $ILP = \pi\mu_0\chi''(\omega)/c$, where $\chi''(\omega)$ is the imaginary part of the complex susceptibility.[57] It has been shown that in case of single-crystalline maghemite, when dispersed in highly viscous media, 14-nm nanoparticles exhibit the maximum heating performance at the frequency and field amplitude of 1000 kHz and 24.8 kA/m.[58] Note that mesoscale ensemble effects such as arrangement,[59] interactions,[60] and alignment[61,62] can lead to deviations. Recently, Lappas et al.[63] observed that the magnetic heating of 10-nm iron oxide nanoparticles is increased by structural point defects inside the crystal lattice (sketch in Figure 1d). In this case the improved heating compared to defect-free particles is explained by increased effective magnetic anisotropy $KV$ thanks to individual point defects.

To further increase magnetic heating, iron oxide nanoparticles with large moments are desired. Large, thermally blocked iron oxide nanoparticles could have a higher heating power than superparamagnetic iron oxide nanoparticles, however only at driving field amplitudes larger than their coercive fields,[64] i.e. much larger than 5 kA/m. For this reason, we propose the use of large and magnetically *soft* iron oxide nanoparticles, so that they have an appreciable response $\chi''(\omega)$ at low field amplitudes.

A possible approach to reduce the effective anisotropy *KV*, and hence increase the initial susceptibility, is the introduction of structural defects. This can be achieved for example by 2D defects in the crystal structure, such as grain boundaries and stacking faults. This approach is well established for iron-based bulk ferromagnets, for which magnetic softening (i.e. decrease of coercivity) can be achieved by decreasing the grain size below the material specific magnetic correlation length.[65–68] Further potential strategies include doping and interfaces, as discussed below.

## 2.2. Defect-rich iron oxide nanoparticles which excel at magnetic hyperthermia

Structural defects, which are routinely found in iron oxide nanoparticles, include vacancies, twinning defects,[69] grain boundaries, and interfaces.[31,63,69–73] Iron oxide nanoparticles synthesized by thermal decomposition method tend to have defected structure and interfaces that can lead to anomalous magnetic properties.[74,75] **Figure 2** shows exemplarily 23 nm iron oxide nanocubes synthesized by thermal decomposition of iron oleate[37,74] Combining high angle annular dark field scanning transmission electron microscopy (HAADF-STEM), X-ray diffraction (XRD), and electron energy loss spectroscopy (EELS), it has been observed that thermal annealing of PEGylated nanocubes at 80°C for 48 h induces $Fe^{2+}$ vacancies in the particle crystal structure. The $Fe^{2+}$ vacancies together with ever remaining FeO-subdomains within the particles led to comparatively small effective magnetic anisotropy *KV*. Remarkably, the thermally treated nanocubes preserved their improved intracellular magnetic hyperthermia, as evaluated by *in-vitro* SAR measurements during interaction and uptake by IGROV-1 breast cancer cells (Figure 2k).

Note that single-phase magnetite nanocubes with same size lose their heating performance nearly completely in viscous media.[76] Thus, these mildly treated nanocubes can be regarded as the prototype for improved magnetic heating by *defect-induced* magnetic softening.

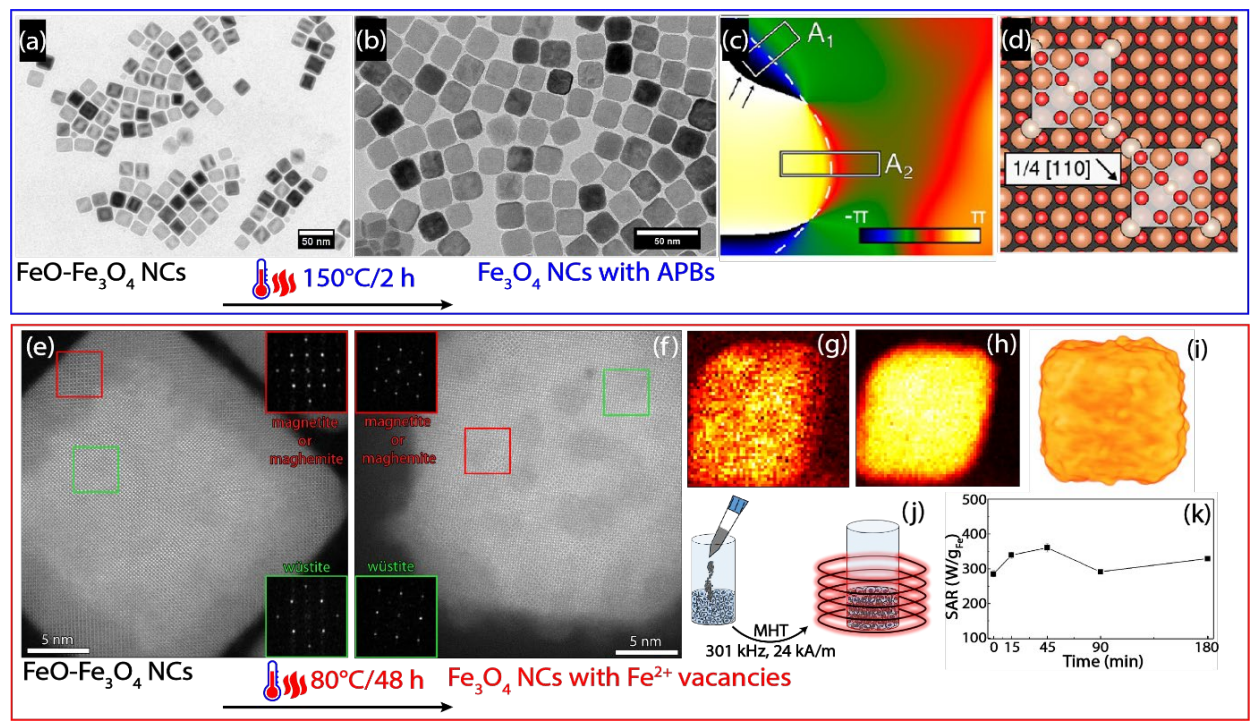

**Figure 2**. **Thermally induced antiphase boundary (APB) defects and cation vacancies in iron oxide nanocubes synthesized via thermal decomposition of iron-oleate**. Low resolution TEM images of (a) as-prepared FeO-$Fe_3O_4$ core-shell nanocubes and (b) thermally annealed nanocubes at 150°C for 2 h. (c) Phase map of {220} spinel-exclusive fringe of a Fourier filtered HRTEM image of a single particle acquired by geometric phase analysis (GPA), indicating an APB with the dashed line, that is formed where two growing magnetite sub-domains coalescent. (d) Scheme of unit cells of two $Fe_3O_4$ domains (white rectangulars) nucleated on an ideal FeO surface that are shifted by 1/4 [110]. Adapted with permission from[74]. Copyright 2013, American Chemical Society. (e) High-resolution HAADF-STEM image of an FeO-$Fe_3O_4$ core-shell nanocube: insets are Fourier analysis diffraction patterns of regions containing FeO and magnetite/maghemite. (f) High-resolution HAADF-STEM image of a nanocube after 48 h annealing at 80°C: insets are Fourier analysis diffraction patterns of regions containing FeO and magnetite/maghemite. (g) $Fe^{2+}$ and (h) $Fe^{3+}$ valency maps of a single thermally treated nanocube obtained by fitting the electron energy loss spectra of each pixel to the reference spectra. (i) Three-dimensional visualization of the individual nanocube shown in (f), which was reconstructed using electron tomography technique. (j) Scheme of evaluating *in-vitro* magnetic hyperthermia performance of thermally treated nanoparticles during interaction and uptake by IGROV-1 breast cancer cells. (k) Evolution

of SAR values as a function of the incubation time were measured at field frequency and amplitude of 301 kHz and 24 kA/m. The SAR is virtually independent of nanoparticle cell internalization and immobilization. Adapted with permission from[37]. Copyright 2018, American Chemical Society.

Flower-shaped iron oxide nanoparticles[32,33] represent a different, prominent example for large, defect-rich iron oxide nanoparticles. As can be seen in **Figure 3a,b,c**, the nanoflowers are 20-100 nm aggregates consisting of small ~ 5-15 nm maghemite crystallites.[77] The small crystallites are separated by grain boundaries, as highlighted by the dashed lines in Figure 3b. For this reason, the nanoflowers are often classified as multicore particles.[48,78] However, because of their dense structure, they can be also regarded as individual magnetic nanoparticles but with a nanocrystalline substructure. Due to exchange interactions between the small crystallites (Figure 3b, dashed lines), the atomic moments within the total particle volume are preferentially magnetized along the same direction, leading to large effective moments at low fields. As a result of the grain boundaries, however, the nanoflowers exhibit a significant internal spin disorder,[35] something which is also observed for nanocrystalline bulk samples of Ni.[79] This spin disorder results in nearly vanishing coercivities,[80] analogous to nanocrystalline bulk ferromagnets.[65–68] Consequently such nanoflowers, and similar dense iron oxide nanoparticle aggregates, excel at magnetic hyperthermia at low field amplitudes.[32–36,80–82] We attribute the increased heating to the magnetic softening, which results in increased $\chi''(\omega)$ at low field amplitudes similar to single-core thermally treated nanocubes discussed before (Figure 2).

An additional remarkable feature of nanoflowers is that their already outstanding heating performance can even be further increased with magnetic interactions at the mesoscale.[34] This can be related to a parallel alignment between neighboring particle moments within dense aggregates[83]. This behavior is in contrast to conventional interacting iron oxide nanoparticle systems (i.e. dense powder samples of single-core particles), in which an antiparallel alignment of neighboring particle moments has been observed[84,85], culminating in reduced susceptibility and decreased heating[86].

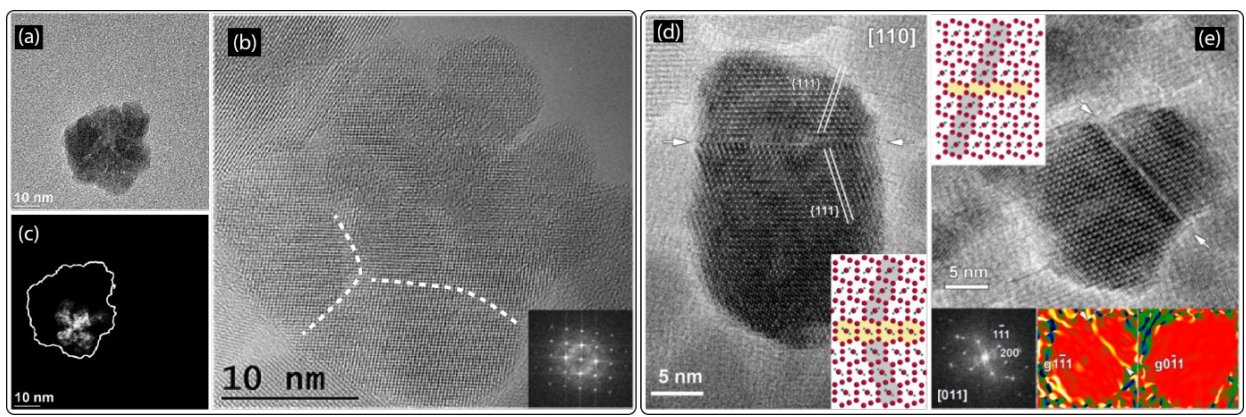

**Figure 3. Grain boundaries and twining defects in iron oxide nanoparticles.** (a) Typical TEM micrograph of an individual nanoflower-shaped iron oxide nanoparticle synthesized via polyol process. (b) HRTEM micrograph of a single flower-like nanoparticle, so-called nanoflower. The dashed lines indicate grain boundaries between domains (c) Dark-field TEM micrograph of an individual particle showing that domains sharing the same crystallographic orientation varies between 5 to 15 nm in size. While each domain is well coherent, domains are oriented differently with respect to each other. (d) HRTEM micrograph of an iron oxide nanoparticle synthesized via hydrothermal method. The nanoparticle shows a twining defect along (111) plane, as indicated by white arrows. Inset shows atoms arrangement along the twining plane that is highlighted in yellow. The flipping of the atomic arrangement of $Fe_3O_4$ around the (111) twining plane is apparent. (e) HRTEM micrograph of a so-called dimer iron oxide nanoparticle synthesized via hydrothermal method, showing a grain boundary defect along (111) crystalline plane. Inset at the top shows atoms arrangement along the grain boundary, that is colored in yellow. Inset at the bottom-left presents Fourier-transform diffraction pattern and insets at the bottom-right show GPA maps of $g_1\bar{1}1$ and $g_0\bar{1}1$ planes. The left side GPA map indicates lattice distortion along the grain boundary plane i.e. (111), as seen by a discontinuity in the color map at the grain boundary. No lattice distortion is observed along out-of-grain boundary planes e.g. (011), as shown in the right side GPA color map. Panels (a,c) adapted with permission from[35]. Copyright 2018, American Chemical Society. Panel (b) adapted with permission from[33]. Copyright 2012, American Chemical Society. Panels (d,e) adapted with permission from[69]. Copyright 2014, American Chemical Society.

As highlighted above, there are multiple iron oxide nanoparticle systems in which structural defects contribute positively to enhancing the performance for biomedical applications that are based on the Néel relaxation mechanism such as magnetic hyperthermia and MPI. In the following, we will review the interrelation of structural and spin disorder in magnetic nanoparticles that is expected to be decisive for their functionality. We will then expand our discussion on the current status regarding defect-engineering in iron oxide nanoparticles, to postulate new pathways for exploiting structural defects in favor of particle performance in biomedical applications.

## 3. Defect-induced spin disorder in iron oxide nanoparticles

In the previous section we introduced several examples of structurally disordered iron oxide nanoparticles that exhibit greatly enhanced performance in biomedical applications, in particular magnetic hyperthermia. Structural deviations from homogeneity can cause atomistic disorder of the magnetic spin ensemble and thus significantly alter the particle properties. In addition, nanoscale surface effects naturally play a decisive role in nanoparticles due to large surface-to-volume-ratio which can also cause localized spin disorder. The spin disorder in magnetic nanoparticles thus needs to be addressed to fully understand the interrelation between defects and disorder and the technological performance of magnetic nanoparticles. Despite recent advances in both microscopy and scattering approaches, resolving the magnetization configuration at the nanoscale[87], including the complex spin configuration within magnetic nanoparticles, remains a key challenge in nanomagnetism.

Whereas long-range order is routinely characterized using diffraction techniques assuming periodic boundary conditions, the local, short range nature of disorder effects requires combination of a variety of techniques on different length scales as well as a model adapted to the length scale that is probed. Structural defects within individual magnetic nanoparticles are typically investigated with high-resolution electron microscopy techniques[71] (**Figure 4a,b**) whereas X-ray techniques such as XRD[63] or X-ray circular dichroism (XMCD)[88] can be used to obtain ensemble-averaged information regarding stoichiometry and crystal structure.

Methodology towards spin disorder includes magnetometry[89], Mössbauer[90] & Raman[91] spectroscopy, nuclear magnetic resonance (NMR)[92], Lorentz transmission electron microscopy (LTEM)[93], high-resolution electron loss spectroscopy (HR-EELS)[70], and magnetic neutron diffraction[94] and small-angle neutron scattering (SANS)[95]. Each of these techniques can provide unique information but only at specific length scales. Therefore, often combinations of various techniques are needed to acquire a self-consistent picture.

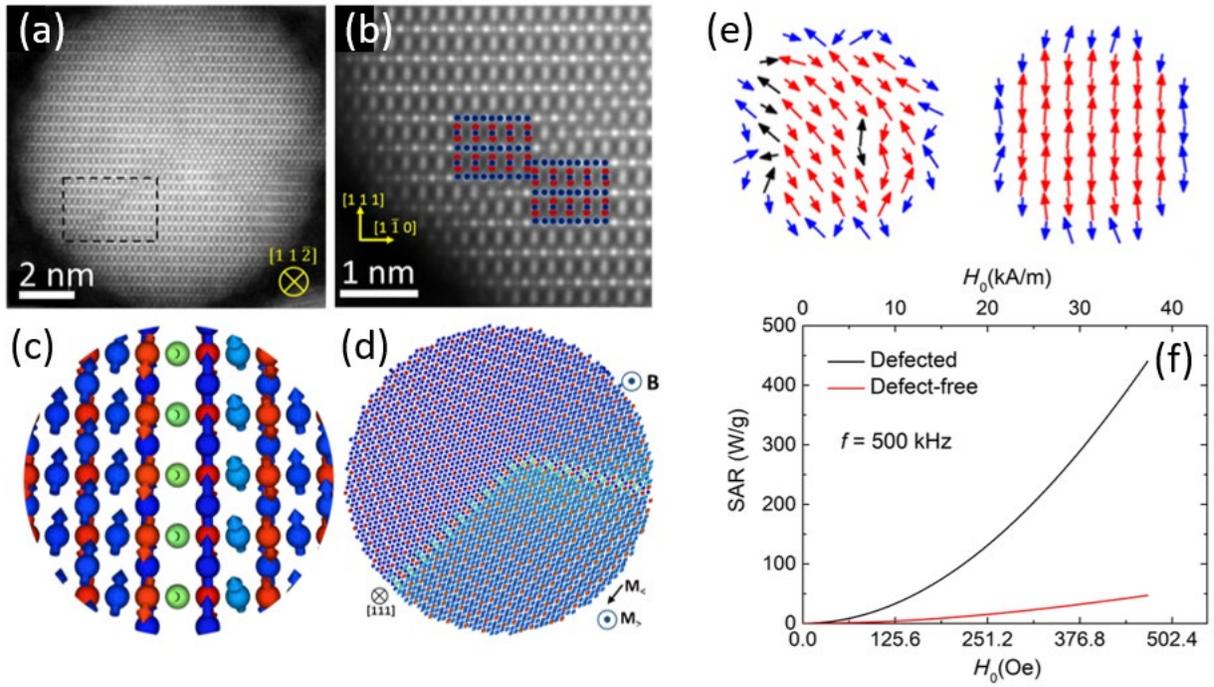

**Figure 4. From structural defects in nanoparticles to spin disorder and enhanced magnetic heating performance**. (a) Anti-phase boundary in a representative iron oxide nanoparticle, atomically resolved using HAADF STEM. (b) magnification of the dashed area indicated in (a). (c,d) Atomistic spin calculations of illustrate the spin misalignment near antiphase boundaries in model nanoparticles. (e) Monte Carlo simulation of the spin ensemble in a defected (left) and a defect-free nanocrystal (right) near field reversal after saturation. (f) Specific absorption rate calculated for a defected and a defect-free nanoparticle. Panels (a-d) and (e,f) reproduced from[71] and[63], respectively.

## 3.1. Spin disorder at different scales

The existence of spin disorder in nanoparticles is typically concluded from low magnetization as compared to the bulk materials[89,96], accompanied by non-saturating magnetic behaviour[97] and exchange biasing[98], all observed using macroscopic magnetization measurements. Microscopic information on spin disorder, i.e. atomic spins deviating from the collinear order of the material, is accessible through the hyperfine structure of in-field Mössbauer spectroscopy[99]. In combination with macroscopic magnetization and X-ray diffraction, the generally assumed nanoscale model of a collinear macrospin surrounded by surface-near spin disorder (Figure 1c) is typically applied[100–103]. Surface spin disorder is understood as a result of broken exchange bonds and low symmetry near the particle surface. Moreover, the correlation of structural and spin disorder is widely accepted. Within the macrospin model, spin disorder is hence mostly attributed to surface effects based on a particle size dependence of macroscopic magnetization properties, or based on a structurally coherent magnetic grain size smaller than the particle itself, as accessible through neutron diffraction[94]. To unambiguously confirm the location of spin disorder near the particle surface, spatially sensitive techniques are required with nanometer resolution.

LTEM provides information on the stray field magnetization with spatial resolution of a few nm and can hence be applied to individual particles with sizes as small as a few nm[93]. Electron holography provides higher spatial resolution and is sensitive to the entire nanoparticle spin configuration and has been applied to the magnetic interparticle coupling in arrangements of nanoparticles[104]. On the individual nanoparticle level, high-resolution magnetic maps of individual Fe nanocubes have been generated to reveal the size-induced transition between vortex (**Figure 5a-c**) and single-domain states (Figure 5d-f) within the nanoparticles[105]. Whereas defect-free nanoparticles have been targeted for such proof-of-principle studies, application of electron holography techniques to the defect-induced spin structure in magnetic nanoparticles will be a challenging and highly interesting endeavor. HR-EELS is both spatially and element-sensitive and has elucidated the site-dependent surface spin canting in cobalt ferrite nanoparticles, revealing surface spin canting for the cobalt sites, but not for the iron sites[70].

Magnetic SANS is sensitive to nanoscale magnetic fluctuations with sub-Å spatial resolution, giving access to the spatially resolved magnetization distribution as well as directionally resolved magnetization correlations in magnetic nanoparticles[95]. Through the spatial sensitivity with length scales of 0.1 – 500 nm, surface spin disorder[106] and correlated surface spin canting[107,108] become accessible. Using magnetic SANS, a significant reduction of surface spin disorder was found upon cooling to low temperatures of 10 K [109]. More recently, a strong field-dependence of surface spin disorder was revealed, expressed by a gradual polarization of initially disordered surface spins even beyond the structurally coherent grain size (Figure 5h)[110]. The field dependence of the spatial distribution of surface spin disorder ultimately gives access to the spatially resolved disorder energy towards the particle surface[110].

Nanoscale surface effects have recently been exploited towards hollow nanoparticles. The high magnetic disorder in these particle shells was attributed to two antiferromagnetically coupled noncollinear structures close to speromagnets based on a broad distribution of hyperfine fields as determined from Mössbauer spectroscopy[111].

### 3.2. Spin disorder – correlated or not?

Depending on surface and magnetocrystalline anisotropies, different types of spin canting towards the particle surface have been suggested by theory, including so-called artichoke, throttled, and hedgehog spin structures[112]. However, experimental evidence for such structures is scarce. The size-dependent spin structures in manganese-zinc ferrite nanoparticles were studied using unpolarized SANS in combination with micromagnetic simulations, revealing increased magnetic inhomogeneity with particle size and correlated, non-collinear spin states occurring for particle sizes beyond 20 nm[113].

Correlated spin canting has been reported for assemblies of smaller, ~10 nm nanoparticles[107,108], without directly discriminating one of the suggested spin structures. In contrast, non-interacting ferrite nanoparticles of similar size reveal random surface spin disorder without a traceable coherent transversal magnetization component[110]. In consequence, a strong impact of interparticle interactions on spin disorder and spin canting is evident and needs to be considered for applications of spin disorder towards magnetic heating.

### 3.3. Spin disorder in the particle interior

Whereas the commonly accepted macrospin model including surface spin disorder is sufficient to explain a large fraction of macroscopic phenomena, the potential existence of atomistic spin disorder in the particle interior (Figure 1d) is often disregarded. The spin disorder distribution within magnetic nanoparticles is only accessible using techniques that are both quantitative and spatially sensitive. Indeed, polarized SANS has revealed significant contributions of spin disorder even in the particle interior[114], expressed as a reduced local magnetization in the particle core[106,109,110]. In several reports, the spin disorder in the particle interior has even been found to dominate surface spin disorder[106,115]. Such a reduced magnetization has been attributed to canting of Fe moments in both tetrahedral and octahedral sites, revealed by a combination of NMR and Mössbauer spectroscopies, and related to reversed moments and frustrated topology[92].

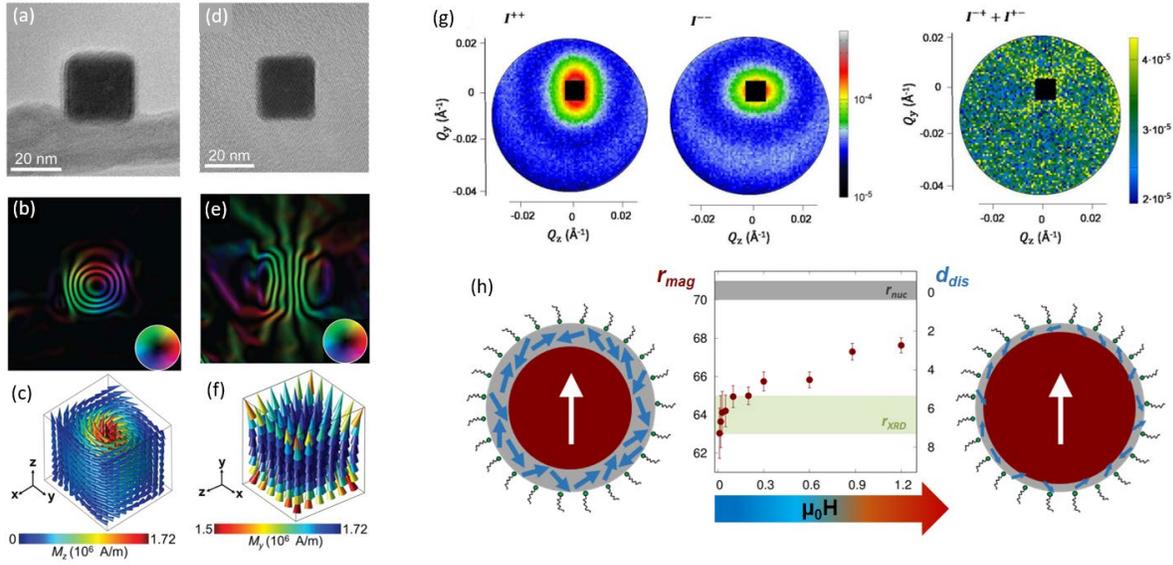

**Figure 5. Nanoscale magnetization in magnetic nanoparticles.** (a-f) Magnetic electron holography of vortex (a-c) and single-domain (d-f) spin configurations in Fe nanoparticles. (a,d) Experimental hologram, (b,e) magnetic induction flux line maps derived from the experimental phase image (inset: color wheel indicating the direction of the magnetic induction). (c,f) Micromagnetic simulation of the 3D magnetization of a $29.5^3$ nm$^3$ and a 24 x 26 x 24 nm$^3$ Fe nanocube in equilibrium state, indicating the vortex and single-domain configurations, respectively. Adapted with permission from[105]. Copyright 2015, American Chemical Society. (g) Magnetic SANS data by cobalt ferrite nanoparticles in applied magnetic field of 1.2 T. The non-spin-flip channels $I^{++}$ and $I^{--}$ give access to the structural morphology and collinear magnetization distribution (*i.e.* the magnetization component parallel to the applied field). The spin-flip channels $I^{+-}$ and $I^{-+}$ provide a combination of collinear magnetization and spin misalignment. (h) Field dependence of the thickness $d_{dis}$ of the surface-near spin disorder shell in ferrite nanoparticles. In increasing applied magnetic field, the size of the collinearly magnetized nanoparticle core $r_{mag}$ overcomes the structurally coherent grain size $r_{XRD}$, leaving a reduced shell of spin disorder. Panels (g,h) adapted with permission from[110].

A clear particle size dependence of spin disorder in iron oxide nanoparticles has been established using Mössbauer spectroscopy, revealing enhanced spin canting in an intermediate particle size of 8-12 nm[31]. Combination of magnetic SANS with nuclear resonance X-ray scattering has further revealed a clear atomic site dependence of the spin misalignment[115]. The significant amount of spin disorder found in the nanoparticle interior underlines that disorder and defects on the atomic scale are highly relevant to understand and control the nanoscale and macroscopic magnetic properties in nanoparticles.

Indeed, structural distortions changing the local coordination geometry in the vicinity of defects have a direct influence on interatomic exchange and hence spin disorder[116]. For ferrite nanoparticles, strong variations of the degree of spinel inversion are commonly observed[117–121]. The synthesis technique has been reported to substantially influence the ferrite nanoparticle magnetization and, hence, also the spin disorder[71], through structural defects such as antiphase boundaries resulting from topotactic oxidation from FeO to maghemite[74] (see Figure 2a-d and 4a-d). Antiphase boundaries have further been found to induce local ferromagnetic coupling and enhance the magnetic properties of magnetic nanoparticles[122]. Local disorder in core-shell $Fe_xO$-$Fe_{3-\delta}O_4$ nanocrystals has recently been discovered by analysis of the atomic pair distribution function and correlated with atomistic spin disorder and enhanced magnetic heating efficiencies[63] (Figure 4e,f). All these findings illustrate that for a comprehensive understanding of the macroscopic performance of magnetic nanoparticles, it is indispensable to consider both structural and spin disorder on a variety of length scales (Figure 1), including the nanoscale magnetization distribution, but also the atomistic scale defect-induced disorder in magnetic nanoparticles.

Having established the potential of structural and magnetic disorder for improved magnetic hyperthermia performance, we next highlight current synthetic approaches on how to deliberately engineer defects and disorder in iron oxide nanoparticles.

## 4. Defect-engineering in iron oxide nanoparticles

### 4.1. Colloidal synthesis: a journey toward disordered iron oxide nanoparticles

The quest to synthesize defect-free and single- crystalline magnetic nanoparticles dates back to almost twenty years ago when the first colloidal thermal decomposition syntheses emerged.

Varieties of organometallic precursors including Fe(C$_6$H$_5$N(O)NO)$_3$[123], Fe(CO)$_5$[124], Fe(acac)$_3$[125], FeO(OH)[126] and iron salts such as Fe(II) acetate[127], have been used for synthesis of iron oxide nanoparticles. Large-scale size and shape-control synthesis of iron oxide nanoparticles have been realized after developing Fe(oleate)$_3$ by Hyeon et al[128]. Despite differing in solvent, capping ligands, and growth temperature, in these non-hydrolytic syntheses, nanoparticles are mostly a product of kinetically driven processes e.g. through adsorption/desorption of capping ligands on certain crystalline facets.[129] Thanks to the kinetic pathways, the formation of uniformly sized and shaped iron oxide nanoparticles often takes no longer than an hour in these methods[130]. However, detailed characterization studies have later revealed that Fe(CO)$_5$, Fe(oleate)$_3$, and Fe(II) acetate lead to FeO-Fe$_3$O$_4$/Fe$_2$O$_3$ and Fe-Fe$_3$O$_4$[131] core-shell iron oxide nanoparticles with degraded magnetization and structural disorders[127,132–135]. Several groups have developed twists in the decomposition chemistry[136] and modified the synthesis involving protective gas[137,138] to eliminate the paramagnetic FeO phase and obtain single-phase Fe$_3$O$_4$ nanoparticles. Further approaches reduce surface spin disorder in superparamagnetic ferrite magnetic nanoparticles by changing the nature of alkaline capping agents[139].

Despite the progress, we know from solid state chemistry and materials science that defect-free crystals are the product of thermodynamically-driven processes, forming at much higher temperatures and longer reaction times than that of colloidal syntheses (i.e. 220-340°C, < 1h). Moreover, recent magnetic hyperthermia studies have strikingly shown that defect-free and highly crystalline single-core iron oxide nanoparticles lose nearly entirely their hyperthermia efficacy in highly viscous media[76], *in-vitro*[49], and intracellularly[37,51]. Di Corato et al. have systematically shown this phenomenon *in-vitro* for multicore and cubic-shape single-core iron oxide nanoparticles[49].

The tendency to compare physicochemical properties of synthetic iron oxide nanoparticles such as saturation magnetization and magnetic anisotropy with the bulk-like properties has thus far distracted us from exploring defects and disorders in favor of biomedical applications. In the following, we will briefly discuss current synthetic approaches for defect- and disorder-engineering in single-phase and interfaces in core-shell iron oxide nanoparticles with particular focus on their impact on magnetic hyperthermia.

## 4.2. Defects and discontinuities induced by growth pathways

Capping ligands are an indispensable part of thermal decomposition syntheses, as they control size and morphology of nanocrystals by influencing the reaction kinetics. Playing with the binding nature and strength between capping agents and nanocrystals has always been an effective strategy to control particle shape and size[140,141]. Recently, capping ligands were exploited to change the internal structure of iron oxide nanoparticles. A prime example is iron oxide nanoflowers (Figure 3a-c), wherein short length polyols such as diethylene glycol allows packing small crystallites into dense disordered structures, being separated by grain boundary defect (dashed lines in Figure 3b)[77,142]. The effect of competition between short and long ligands on the particle structure has recently been shown in zinc ferrite nanoparticles. Increasing the ratio of acetylacetone to oleic acid ligands results in assembly of small magnetic zinc ferrite subdomains into 100 nm cubic-particles, due to less bulkiness and higher mobility of acetylacetone ligands as compared to oleic acid[143].

Hydrothermally synthesized iron oxide nanoparticles showed to possess twining and stacking faults defects (Figure 3d,e[69]. Formation of twining defects and stacking faults along (111) crystalline planes was attributed to the oriented-attachment crystal growth mechanism. A growth mechanism similar to the nanoflowers was proposed for these defected iron oxide nanoparticles. The fusion of initial $Fe_3O_4$ crystallites, that is promoted by desorption of capping ligands, into a single particle accounts for twining and stacking fault defects. Interestingly, none of these defects were observed when a similar synthesis mixture was treated by co-precipitation synthesis method. It was therefore discussed that an intensive activity and mobility of the capping ligands on crystallites, given to the synthesis reaction by high temperature hydrothermal reaction conditions, is required to form defect-rich iron oxide nanoparticles. Remarkably, it was demonstrated that iron oxide nanoparticles with stacking fault internal defects have a higher $M_s$ and SAR than defect-free iron oxide nanoparticles[69].

**4.3. Defects and vacancies induced by oxidation**

Most iron oxide nanoparticles synthesized via thermal decomposition syntheses have to some extent defects, disorders, and sub-domains. $FeO$-$Fe_3O_4$ core-shell nanoparticles, synthesized via the thermal decomposition of $Fe(oleate)_3$ and $Fe(CO)_5$ are the most heavily characterized iron oxide nanoparticle systems due to the synthesis robustness and particle size uniformity[144]. Studies on post synthesis thermal treatment of these iron oxide nanoparticles toward single-phase iron oxide have led to new insights into oxidation-induced defects. It has been shown that the nucleation and growth of $Fe_3O_4$ on the FeO phase (i.e. the topotaxial oxidation from FeO to magnetite and maghemite) lead to formation of antiphase boundary (APB) defects at the interface between growing magnetite domains. Interestingly, APBs remained after four hours of thermal treatment at 150°C (Figure 2a-d)[74]. The oxidation of core-shell iron oxide nanoparticles, synthesized by decomposition of $Fe(CO)_5$, via mild cyclic magnetic stimulations has also resulted in the APBs[145]. A step toward exploiting structural defects for biomedical applications was taken by mild thermal treatment of $FeO$-$Fe_3O_4$ nanoparticles after being capped with catechol-terminated functional

poly(ethylene glycol) polymers. A 48 hours thermal treatment at 80°C induced $Fe^{2+}$ vacancies in the $Fe_3O_4$ phase of core-shell iron oxide nanoparticles, which result in the magnetic softening and thus the domination of the Néel relaxation. As discussed in the previous section, such a magnetic softening lead to improved magnetic hyperthermia performance of 23 nm cubic-shape particles compared to single-phase counterparts[37] (Figure 2e-k).

## 4.4. Structural defects associated with interfaces and doping

Interfaces and doping are additional promising sources of breaking composition continuity and inducing atomic scale exchange interactions in magnetic nanoparticles. The magnetic anisotropy can be tuned by magnetic coupling in bi-magnetic hard-soft and soft-hard core-shell magnetic nanoparticles such as $CoFe_2O_4$-$MnFe_2O_4$[146] and $Fe_3O_4$-$CoFe_2O_4$[147]. Recently, trimagnetic $Fe_3O_4$-$CoO$-$Fe_3O_4$ core-shell-shell nanoparticles have been developed via seed-mediated thermal decomposition synthesis. Hard-soft exchange couplings at $Fe_3O_4$-Co doped ferrite and Co-doped ferrite-$Fe_3O_4$ interfaces largely influenced the superparamagnetic blocking temperature and magnetization hysteresis loops[148].

An emerging approach to furthermore enhance magnetic exchange couplings is to tune the shell thickness. It was recently shown that forming an ultrathin shell of soft magnets (e.g. $Fe_3O_4$ and $MnFe_2O_4$) on $CoFe_2O_4$ nanoparticles leads to increased magnetic anisotropy[149,150]. The shell thickness turned to be critical to observe a so-called enhanced spin canting phenomenon. Inserting a paramagnetic FeO interlayer spacer between Fe core and iron oxide shell in layered iron oxide nanoparticles was led to magnetostatic interactions between the core and the shell and enhanced magnetic hyperthermia performance[73]. Despite being a promising approach, one has to bear in mind that synthesis of multi layered nanoparticles is much more challenging than those Iron oxide nanoparticles formed via one-pot reactions.

Breaking the cation ordering by substituting magnetic[151] and non-magnetic ions into the iron oxide crystal structure has emerged as a strong tool to tune magnetic properties in the past years[152,153]. For spinel ferrite nanoparticles, the cation distribution between A and B sites is decisive for the macroscopic magnetization, and strong variations of the inversion degree have been observed as nanoparticles often represent a non-equilibrium cation distribution[117–119]. Zinc and manganese are the most commonly substituted cations into iron oxide nanoparticles[154]. Despite being non-magnetic, $Zn^{2+}$ replaces $Fe^{3+}$ in tetrahedral crystalline sites and thus reduces antiferromagnetic couplings between $Fe^{3+}$ in tetrahedral and octahedral sites. Depending on the ionic size and concentration of the substituting cation, lattice strain was induced in cobalt ferrite nanoparticles.[155,156] Such approach appears promising to tailor the lattice strain and in result control the magnetization of iron oxide nanoparticles.

## 5. Conclusions and perspectives

The strong research interest on iron oxide nanoparticles is particularly driven by their unique potential for a wide range of biomedical applications. To optimize their magnetic properties, historically, the usual strategy has been to strive for the synthesis of defect-free single crystalline nanoparticles. The particle characteristics have been then primarily varied by their morphological features such as size and shape. In this progress report we have compiled recent findings indicating an alternative approach for improving magnetic particle properties. Inspired by other research fields such a semiconductors and perovskite solar cells, we here attempt to redirect the attention of the magnetic nanoparticle community toward exploiting defect-engineering in iron oxide nanoparticles to enhance and adapt particle magnetic properties for *in-vivo* magnetic hyperthermia. Recent experimental and computational studies emphasize a clear benefit of structural defects and associated magnetic spin disorder in various biomedical applications. Defect-rich iron oxide nanoparticles appear particularly promising candidates for intracellular magnetic hyperthermia and magnetic-heating-triggered drug delivery, wherein transducing heat based on the Néel magnetic relaxation mechanism is crucial. Sharing similar nanomagnetism principles to magnetic hyperthermia[157], magnetic particle imaging (MPI), an emerging molecular imaging modality, will largely benefit from defect-engineering in iron oxide nanoparticles. The research on developing iron oxide based tracers for MPI is growing rapidly owing to outstanding features of MPI compared to MRI such as high sensitivity, deep tissue imaging, and capability of live imaging[158]. It is well known that the MPI imaging technique works based on the Néel relaxation mechanism. Therefore, iron oxide nanoparticles with the characteristics highlighted here for defect-rich iron oxide nanoparticles, that is large magnetic susceptibility and low magnetic anisotropy, are well suited for MPI. This correlation was verified by several studies which conclude that nanoflowers are excellent candidates for MPI[38,39].

We have here summarized studies on synthesis of defect-rich iron oxide nanoparticles, yet our focus is on those studies in which defects and disorder have been mainly included deliberately. We have highlighted few one-pot synthesis reactions and post synthesis thermal treatments in organic solvents and aqueous media that promote structural defects and compositional discontinuities in iron oxide nanoparticles. Anti-phase boundaries, stacking faults, grain boundaries, and twining defects have been induced in iron oxide nanoparticles.

Looking into the correlation between structural and magnetic lattice, we have realized that one important aspect that is often overlooked is the different characteristics of spin disorder on nanoscale and atomistic length scales. The magnetic structure of iron oxide nanoparticles is directly determined by their structural features, implying that structural defects lead to disordered magnetic lattice and eventual modification of the magnetic properties. Despite the progress in atomistic analyses, the precise characterization of spin disorder within the particles remains one of the key challenges in nanomagnetism. However, it is crucial to distinguish between surface-induced nanoscale spin disorder and atomistic, structurally induced intraparticle spin disorder when assessing macroscopic magnetic properties and hence technological performance of magnetic nanoparticles.

Admittedly, the programmed defect inclusion in iron oxide nanoparticles is yet in its infant stage, when compared to other fields such as semiconductor technology. However, the number of research articles on revealing and harnessing positive contribution of defects is rapidly growing. In the following, we add our perspectives and ideas on how to systematically engineer defects into the crystal structure of iron oxide nanoparticles.

One of the potential approaches to synthesize iron oxide nanoparticles with grain boundaries like nanoflowers discussed here, is to use a combination of short and long capping ligands with different binding affinity to iron. A combination of short phosphonate-based and oleic acid ligands in thermal decomposition synthesis could lead to the formation of iron oxide nanoparticles with grain boundaries via competitive ligand binding to different positions on particles. The nature of binding between capping ligands and particles is yet largely unknown and thus more elucidations in this regard will facilitate designing disorder in iron oxide nanoparticles.

We envision a great opportunity in combining crystalline disorder and *in-vivo* clearance of magnetic nanoparticles in developing defect-rich nano-heaters and MPI tracers. We believe a novel synthetic approach could be to fabricate nanoparticle organic frameworks (NOFs), wherein nanoparticles are assembled together with responsive molecules such as peptides and short oligonucleotides. Controlling size and polydispersity of these assemblies, NOFs behave similar to nanoflowers and excel at intracellular magnetic hyperthermia. The unique feature of NOFs, non-existent in nanoflowers, is the capability of being dissociated to small building blocks that can be easily cleared from the body. NOFs will be particularly promising for magnetic monitoring of drug release, wherein a frequent injection of nanocarriers e.g. once a week and thus a rapid clearance of previously administrated nanocarriers is needed to not interfere with newly coming ones in terms of magnetic signal.

A further strategy is doping-induced-defects by replacing iron with magnetic and nonmagnetic cations such as Cu, Co, Zn, or Gd[159,160]. Doping in iron oxide nanoparticles has traditionally been mainly used to reduce compensating magnetic couplings and improve magnetization, as in the case of $Zn^{2+}$ in iron oxide nanoparticles. We envision the opportunity of inducing other defects and thus tailoring magnetic anisotropy of nanoparticles by defect-induced distortion of the magnetic lattice. Doping appeared as a powerful tool to improve electrical properties of semiconductors[6] and modify photonic properties of quantum dots[161]. Studies on doping-induced-defects in iron oxide nanoparticles seem to lack behind other technological fields, which call for employing a combination of atomistic analyses to unveil defects and their impacts on magnetic properties in near future to come.

To sum up, we advocate our review will stimulate further research and development on novel synthetic routes for defect-engineering as well as selecting appropriate characterization methods depending on the type of defect in magnetic nanoparticles. Eliminating the negative connotation associated with defects and disorder in magnetic nanoparticles, our report will set the scene for embracing the structural and magnetic lattice disorders in magnetic nanoparticles to further advance *in-vivo* biomedical applications.

**Acknowledgements**
S.D. acknowledges financial support from the German Research Foundation (DFG Grant DI 1788/2-1). A.L. acknowledges the Alexander von Humboldt Foundation for postdoctoral research fellowship. P. B. acknowledges financial support from the National Research Fund of Luxembourg (CORE SANS4NCC grant).
All three authors contributed equally to the manuscript.